\documentclass[floatfix,groupedaddress,superscriptaddress,aps, amsmath,amssymb, pra,longbibliography, twocolumn,a4]{revtex4-1}

\usepackage{graphicx,float}
\usepackage{subfigure}
\usepackage{dcolumn}
\usepackage{bm}
\usepackage{mathrsfs}
\usepackage{txfonts}
\usepackage{CJK}
\usepackage{epsfig}
\usepackage{lipsum}
\usepackage{color}
\usepackage{placeins}
\usepackage[colorlinks]{hyperref}
\usepackage{xcolor}

\makeatletter
\emergencystretch=\maxdimen
\newcommand{\rmnum}[1]{\romannumeral #1}
\newcommand{\Rmnum}[1]{\expandafter\@slowromancap\romannumeral #1@}
\makeatother

\hypersetup{    plainpages=true,
    breaklinks=true,    hypertexnames=false,    pageanchor=true,
    colorlinks=true,
    linkcolor={blue},
    citecolor={blue},
    urlcolor={blue},
    anchorcolor={black}
}

\begin{document}
\title{Classical route to quantum chaotic motions}

\author{Nan Yang}\email{nyang.hust@gmail.com}
\affiliation{Theoretical Quantum Physics Laboratory, RIKEN,
Saitama, 351-0198, Japan} \affiliation{Center for Quantum Science and Engineering, and Department of Physics, Stevens Institute of Technology, Hoboken, New Jersey 07030, USA}

\author{Xuedong Hu}
\affiliation{Department of Physics, University at Buffalo, SUNY, Buffalo, New York 14260-1500, USA} \affiliation{Theoretical Quantum Physics Laboratory, RIKEN,
Saitama, 351-0198, Japan}

\author{Yong-Chun Liu}
\affiliation{State Key Laboratory of Low-Dimensional Quantum Physics, Department of Physics, Frontier Science Center for Quantum Information, Collaborative Innovation Center of Quantum Matter, Tsinghua University, Beijing, 100084, China}

\author{Ting Yu}
\affiliation{Center for Quantum Science and Engineering, and Department of Physics, Stevens Institute of Technology, Hoboken, New Jersey 07030, USA}

\author{Franco Nori}
\affiliation{Theoretical Quantum Physics Laboratory, RIKEN,
Saitama, 351-0198, Japan} \affiliation{Physics Department, The University
of Michigan, Ann Arbor, Michigan 48109-1040, USA}
\date{\today}

\begin{abstract}
We extract the information of a quantum motion and decode it into a certain orbit via a single measurable quantity. Such that a quantum chaotic system can be reconstructed as a chaotic attractor.
Two configurations for reconstructing this certain orbit are illustrated, which interpret quantum chaotic motions from the perspectives of probabilistic nature and the uncertainty principle, respectively. We further present a strategy to import classical chaos to a quantum system, revealing {a} connection between the classical and quantum worlds.
\end{abstract}

\maketitle

\textit{Introduction} --- Chaotic motion is known for initial condition sensitivity and ergodicity.
Its characteristics were gradually revealed over the past century, leading to the emergence of chaos theory~\cite{strogatz2001,Moon1992,Ott2002}. For classical systems, chaos theory provides powerful tools to analyze and understand chaotic motions~\cite{Grebogi1997,Field1995,Field1997}.

The study of quantum chaos has attracted wide attention over the past few decades.  On the surface, quantum chaos seems paradoxical: while chaos theory is built on orbits and deterministic motion, physical observables of a quantum system are described probabilistically. It is thus quite natural to assume that the existing chaos theory, developed for classical systems, cannot be applied to quantum systems directly. Nevertheless, explorations on
quantum chaotic dynamics and the quantum origins of classical chaos have touched many branches of physics, from atomic physics~\cite{ruder1994atoms,Blumel2005,saif2005classical}, to quantum transport~\cite{Sebastian2007}, to complex spectra~\cite{Izrailev1990}, etc.  Some endeavors started from finding the semiclassical orbits of quantum objects~\cite{Nakamura1993,Heller1984,Heller2018,semiclassical_1,semiclassical_2,RS1,RS2,RS3}, {while others sought the statistical and spectral features leading to chaos}~\cite{TingYu1999,Naghiloo2017,Mourik2018,RS1,RS2,RS3,random_matrix1,random_matrix2,Haake1991,
energy_level_1,kick_rotor1,Slomczynski1994,Zurek1995QuantumCA,Chirikov1995,
kowalewska2008wigner,geisel2000quantum}, such as complex spectra~\cite{RS1,RS2,RS3,random_matrix1,random_matrix2,Haake1991,energy_level_1} and ergodic phenomena~\cite{kick_rotor1,Slomczynski1994,Zurek1995QuantumCA,Chirikov1995}.
Even though great progress has been made, certain fundamental questions remain open, such as how to define chaos in a quantum system, and whether the linear Schr\"{o}dinger equation would rule out chaos~\cite{kick_rotor1}.

While a quantum system in a superposition state is probabilistic when measured, its energy eigenstates evolve completely deterministically. Furthermore, the Schr\"{o}dinger equation can also be nonlinear if a quantum system is coupled to a nonlinear classical system. It is thus conceivable that chaotic features could be directly observable in a quantum system.
%
In this letter, we extract the dynamics of a quantum chaotic system via a single measurable quantity, and {then} decode it into a deterministic chaotic attractor. Therefore,
the time evolution of a quantum object can be tracked by a certain deterministic orbit in phase space. Moreover, we "import" classical chaos into a quantum object, breaking the rule of linearity in quantum systems.

For concreteness, we restrict our discussions to finite-dimensional quantum systems. Generally, chaos theory built on deterministic orbits are determined by state variables and deterministic functions.
To be consistent with chaos theory, we propose two configurations (I and II), for each one that a quantum system is described by state variables and a deterministic function. Physically, the state variables in configurations I and II quantitatively describe quantum uncertainty and probability distributions, respectively.

\begin{center}
\begin{figure}[]
\centering
\includegraphics[width=3.5 in]{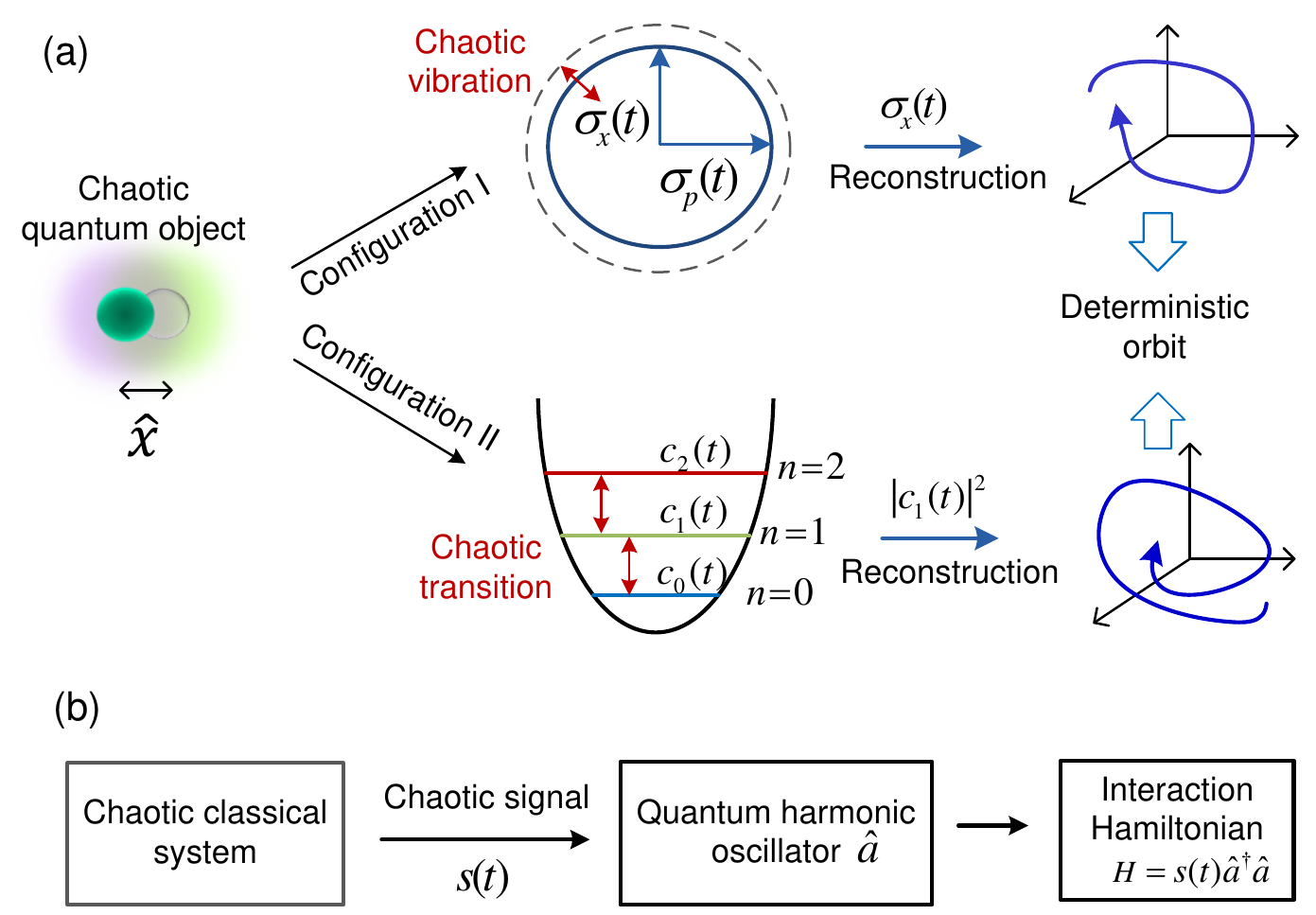}
\caption{(color online) (a) Schematic of constructing a certain deterministic orbit from a quantum chaotic motion. The information of chaos is encoded in the standard derivation $\sigma_x$ in Configuration I or the norm squared probability amplitude $|c_n|^2$ of the \textit{n}-th energy level $|n\rangle$ ($n$=1,2,...) in Configuration II. (b) Schematic for importing classical chaos into a quantum system. Here, the classical chaotic signal $s(t)$ acts on the resonant frequency term of the quantum system.}
\label{fig01}
\end{figure}
\end{center}

\begin{itemize}
\item
Configuration I: the state variables are the standard deviations $(\sigma_{x}, \sigma_{p})$ of the quantum position and momentum ($\hat{x}$, $\hat{p}$): $\sigma_{x}=\sqrt{\langle \hat{x}^2 \rangle-\langle \hat{x} \rangle^2}$ and $\sigma_{p}=\sqrt{\langle \hat{p}^2 \rangle-\langle \hat{p} \rangle^2}$. Their deterministic function is derived from the master equation, e.g., $d\langle \hat{x}^2 \rangle/dt={\rm Tr}{(\dot{\rho} \hat{x}^2)}$ and $d\langle \hat{x} \rangle/dt={\rm Tr}{(\dot{\rho} \hat{x})}$, where $\rho$ is the density matrix.
\item
Configuration II: the state variables are norm squared probability amplitudes ($|c(y_{1})|^2$, $|c(y_{2})|^2$, ...$|c(y_{n})|^2$) of possible quantum states ($|y_{1}\rangle$, $|y_{2}\rangle$, ...$|y_{n}\rangle$). While, their deterministic function is obtained from the master equation since
($|c(y_{1})|^2$, $|c(y_{2})|^2$, ...$|c(y_{n})|^2$) are the coefficients of
the diagonal elements of the density matrix $\rho$.
\end{itemize}
Importantly, we have quantified the evolution of a quantum object by variables instead of wave functions or operators. As shown in Fig.~\ref{fig01}(a), chaos in the sense of Configuration I or II can thus be interpreted as the chaotic vibration of the standard deviation or the chaotic transitions among possible states in quantum systems. The next step is how to construct deterministic orbits from them. According to Takens theorem~\cite{Takens}, even a single measurable quantity preserves enough information to reconstruct the orbit of an unknown chaotic system if it participates in the system evolution. This is the so-called time-delayed coordinates phase-space reconstruction. We can find such a measurable quantity in configuration I (II), e.g., $\sigma_{x}(t)$ [$|c(y_{1})|^2$], and thus obtain a certain deterministic orbit from a probabilistic quantum system.

Hereafter, we define a certain deterministic orbit reconstructed from a quantum system, as a "generalized quantum orbit". It indicates the motion of a quantum object, e.g., the occurrence of a chaotic attractor implies that
the quantum system is in a chaotic regime.

Before proceeding to the details of these configurations, it is necessary to discuss how chaos occurs in a quantum system.
In this work, we propose a scenario where chaos in quantum systems is imported from classical chaos. As shown in Fig.~\ref{fig01} (b): a classical chaotic signal $s(t)$ acts on the frequency term of a quantum system. With this classical-quantum coupling, a quantum system originally in a linear regime can be converted to a chaotic state.

To study quantum chaotic systems with the above proposals, we present an optomechanical setup for either configuration~\cite{Aspelmeyer2014,Jiang2017,Sciamanna2016,Monifi2016,Bakemeier2015,Buters2015,Carmon2007,Larson2011,Lee2009,Lv2015,Ma2014,Marino2013,
Navarro-Urrios2017,Piazza2015,Sun2014,Suzuki2015,Walter2016,WangGuanglei2014,
WuJiagui2017,YangNan2015,ZhangK2010,ZhuGL2019,Ozdemir2019}.
Here, each setup consists of both classical and quantum components, and chaos is imported from the classical to the quantum parts.
These quantum chaotic motions can be visualized and verified by the corresponding generalized quantum orbits.

\begin{center}
\begin{figure}[t]
\centering
\includegraphics[width=3.5in]{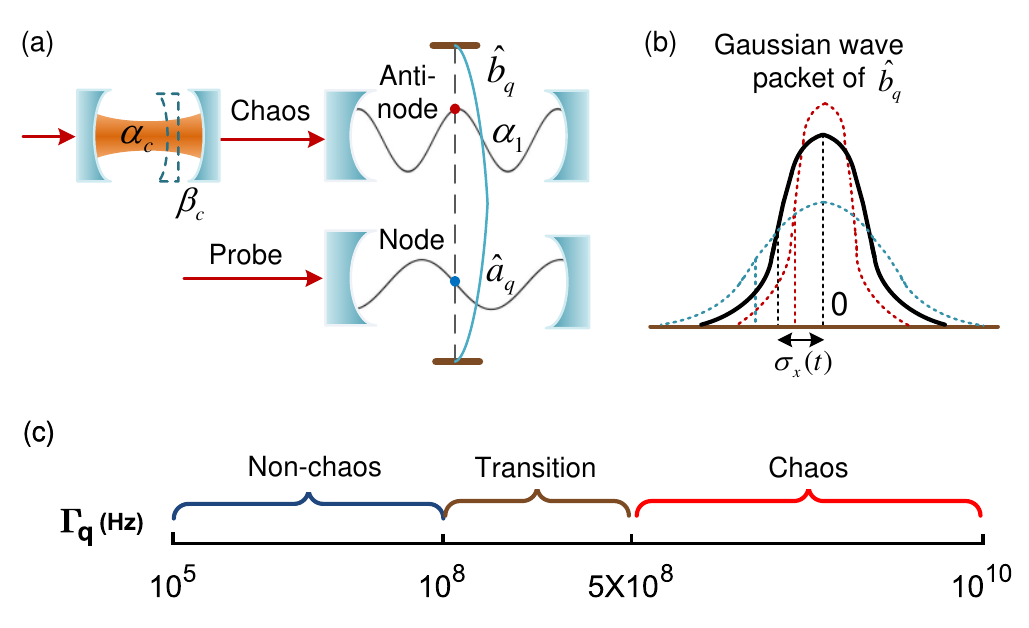}
\caption{(color online) (a) An optomechanical setup for generating chaos in a quantum harmonic oscillator in configuration I. (b) The quantum wave packet determined by $(\sigma_{x},\sigma_{p})$ varies under the influence of the classical chaotic field $\alpha_1$. (c) Our numerical simulations indicate that the transfer of chaos mainly depends on the quantum mechanical damping rate $\Gamma_q$.}
\label{fig02}
\end{figure}
\end{center}

\textit{Configuration I.}
This setup is shown in Fig.~\ref{fig02}(a).
Our {goals} are to import classical chaos generated in a classical optomechanical resonator ($\alpha_c$, $\beta_c$) into a quantum mechanical resonator $\hat{b}_q$ and study its quantum chaotic motion by the standard deviations $(\sigma_{x},\sigma_{p})$. To achieve the former one, a quantum mechanical membrane $\hat{b}_q$ is placed at an anti-node of the classical cavity $\hat{a}_1$ and at a node of the quantum cavity $\hat{a}_q$, which is quadratically coupled to $\alpha_1$ and linearly coupled to $\hat{a}_q$, respectively. The interaction Hamiltonians read
\begin{equation}\label{Interaction}
H_{1, \rm int}={g_1} |\alpha_1|^2 \hat{b}_q^{\dag}\hat{b}_q, \,\,\,\,\,\,\, H_{q, \rm int}=g_q \hat{a}_q^{\dag}\hat{a}_q (\hat{b}_q + \hat{b}^{\dag}_q),
\end{equation}
where $g_1$ ($g_q$) is the coupling strength between the cavity mode $\alpha_1$ ($\hat{a}_q$) and the
mechanical mode $\hat{b}_q$. In this arrangement, the classical cavity $\alpha_1$ is used to imports chaos, while the quantum cavity $\hat{a}_q$ inputs quantum fields into the quantum mechanical mode $\hat{b}_q$.

%

We first focus on the classical parts ($\alpha_c$, $\beta_c$, $\alpha_1$). Here the cavity mode $\alpha_1$ is prepared to a chaotic state by coupling to a chaotic optomechanical resonator ($\alpha_c$, $\beta_c$).
Their equations of motion are given by
\begin{subequations}\label{classical_eq_c2}
\begin{align}
\dot{\alpha}_1 &={-i\Delta_1 \alpha_1} - \frac{\gamma_1}{2}
\alpha_1
-\sqrt{\gamma_1 \gamma_c}\,{\alpha}_c\;,\\
\dot{\alpha}_c &=-i\Delta_c \alpha_c - \frac{\gamma_c}{2}
\alpha_c-i g_c \alpha_c\, (\beta_c + \beta_c^*)
+\varepsilon_c\;,\\
\dot{\beta}_c &= \left(-i \Omega_c - \frac{\Gamma_c}{2}\right)\, \beta_c - i g_c |\alpha_c|^2\;.
\end{align}
\end{subequations}
where $\Delta_1$ ($\Delta_c$), $\gamma_1$ ($\gamma_c$), and $\varepsilon_1$ ($\varepsilon_c$) denote the detuning, the damping rate, and the driving strength of the cavity mode $\alpha_1$ ($\alpha_c$). While, $\Omega_c$ and $\Gamma_c$ are the resonance frequency and the damping rate of the mechanical mode $\beta_c$, and $g_c$ is the coupling strength between $\alpha_c$ and $\beta_c$.
In this model, the optical field $\alpha_1(t)$ input into the quantum part links both the classical and quantum components together.


\begin{center}
\begin{figure*}[t]
\centering
\includegraphics[width=7in]{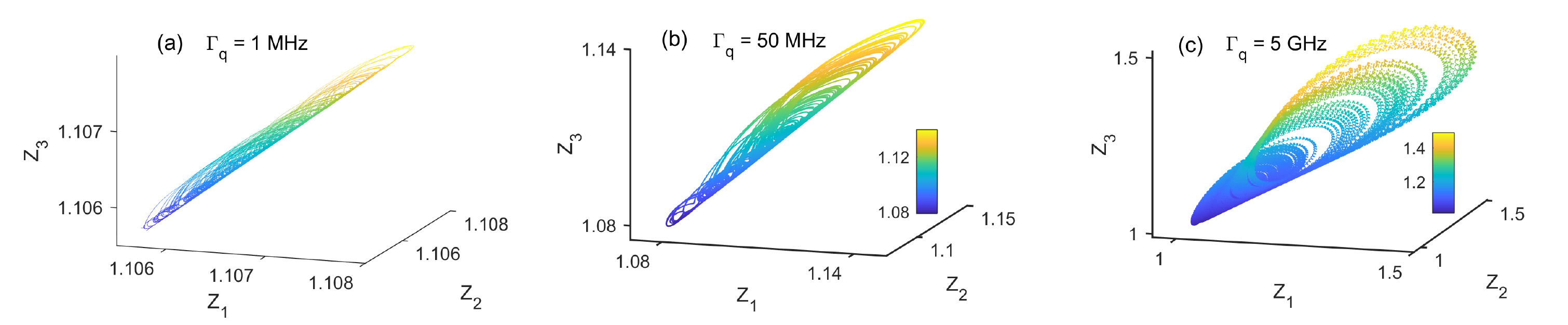}
\caption{(color online) {Generalized quantum orbits reconstructed from the measurable quantity $\sigma_x(t)$ for different mechanical damping rates: (a) $\Gamma_q=1$~{\rm MHz}, (b) $\Gamma_q=50$~{\rm MHz}, and (c) $\Gamma_q=5$ ~{\rm GHz}. Here, $\sigma_x(t)$ is embedded into a 4D phase space with coordinates ($z_1$, $z_2$, $z_3$, $z_4$), where the fourth coordinate $z_4$ is represented by scaled colors.
The parameters chosen here are:
$\Delta_c/\Omega_c=-1$, $\gamma_c/\Omega_c=1$, $g_c/\Omega_c=10^{-3}$,
$\varepsilon_c/\Omega_c=433$,
$\Gamma_c/\Omega_c=2.8$, $\Omega_c/2\pi=0.01~\rm GHz$, $\Delta_1/\Omega_q=10^4$, $\gamma_1/\Omega_q=6$, $\varepsilon_1/\Omega_c=0$, $\Delta_q/\Omega_q=-2$, $\gamma_q/\Omega_q=1$, $g_1/\Omega_q=2$, $G_q/\Omega_q=10^{-4}$, $\Omega_q/2\pi=0.1~\rm GHz$, and
$T=0.002~\rm K$}.}
\label{fig03}
\end{figure*}
\end{center}

Then, we now turn to the quantum parts, i.e., the quantum optical and mechanical modes ($\hat{a}_q$ and $\hat{b}_q$).
We divide $\hat{a}_q$ and $\hat{b}_q$ into the classical and the quantum components: $\hat{a}_q = \langle \hat{a}_q \rangle + \hat{\tilde{a}}_q$ and $\hat{b}_q=\langle \hat{b}_q \rangle + \hat{\tilde{b}}_q$.
Hence, $\sigma_{x}$ can then be rewritten as
\begin{equation}\label{xp4}
\sigma_{x}=\sqrt{\frac{1}{2} + \langle \hat{\tilde{b}}_q^\dag \hat{\tilde{b}}_q\rangle + {\rm Re}[\langle \hat{\tilde{b}}_q^2\rangle]}.
\end{equation}
It can be seen that $\sigma_{x}$ mainly depends on $\langle \hat{\tilde{b}}_q^\dag \hat{\tilde{b}}_q \rangle$, as $\langle \hat{\tilde{b}}_q^2\rangle$ is comparatively small. Here, the equation of motion of $\langle \hat{\tilde{b}}_q^\dag \hat{\tilde{b}}_q\rangle$ takes the form (Supplementary \rmnum{1})
\begin{equation}\label{Linear_eq}
\begin{split}
{d\langle \hat{\tilde{b}}_q^\dag \hat{\tilde{b}}_q \rangle}/dt =& -i G_q (-\langle \hat{\tilde{a}}_q^\dag \hat{\tilde{b}}_q \rangle+\langle \hat{\tilde{a}}_q^\dag \hat{\tilde{b}}_q \rangle^*+\langle \hat{\tilde{a}}_q \hat{\tilde{b}}_q \rangle^*-\langle \hat{\tilde{a}}_q \hat{\tilde{b}}_q \rangle) \\
&-\Gamma_q \langle \hat{\tilde{b}}_q^\dag \hat{\tilde{b}}_q \rangle+\Gamma_q n_{\rm th}[\alpha_1(t)],
\end{split}
\end{equation}
where $\Gamma_q$ is the mechanical damping rate, and {$n_{\rm th}[\alpha_1(t)]=K_{\rm B}T/\hbar[\Omega_q +  g_1 |\alpha_1(t)|^2]$} is the mean thermal phonon excitation number when the environmental temperature is $T$. Here, $G_q=g_q|\langle\hat{a}_q\rangle|$ is the linearized coupling strength. While, the equations for $\langle \hat{\tilde{a}}_q^\dag \hat{\tilde{b}}_q \rangle$ and $\langle \hat{\tilde{a}}_q \hat{\tilde{b}}_q \rangle$, together with Eq.~(\ref{Linear_eq}), are obtained from the master equation of the linearized system (Supplementary \rmnum{1}), which can be also found in Ref.~\cite{Wilson2008,Yong-Chun2013}.

From Eqs. (\ref{xp4}) and (\ref{Linear_eq}), we find that $\sigma_{x}$ is a function of the classical input $\alpha_1(t)$. Here, the classical input $\alpha_1(t)$ is the origin of nonlinearity in the quantum mechanical resonator $\hat{b}_q$. Without $\alpha_1(t)$, Eq.~(\ref{Linear_eq}) degenerates to a linear system and $\sigma_{x}(t)$ finally converges to a fixed value.

When $\alpha_1(t)$ is chaotic, to check if the quantum mechanical resonator $\hat{b}_q$ is also driven to a chaotic state, we study its generalized quantum orbit reconstructed from the standard deviation $\sigma_{x}(t)$. Among all the parameters, we find that the mechanical damping rate $\Gamma_q$ is the crucial one for this chaos transfer [Fig.~\ref{fig02}(c) and Supplementary video \rmnum{1}]. Chaos can only be imported into the quantum mechanical resonator $\hat{b}_q$ when $\Gamma_q > 0.1$~{\rm GHz}.
Figure~\ref{fig03} illustrates the corresponding generalized quantum orbits of $\hat{b}_q$ for different $\Gamma_q$. It can be seen that a chaotic attractor emerges in the phase space [Fig.~\ref{fig03}(c)] when $\Gamma_q$ is increased to $5$~{\rm GHz}. This chaos is also indicated by the positive largest Lyapunov exponent (+0.12).
Here, the generalized quantum orbits are obtained by embedding $N$ time-delayed coordinates $\sigma_x(\tau)$, $\sigma_x(2\tau)$, ..., $\sigma_x(N\tau)$ into a 4D phase space ($z_1$, $z_2$, $z_3$, $z_4$), where $\tau=0.3$~{\rm ns}. Experimentally, $\sigma_x$ is a measurable quantity and can be read out by existing technologies~\cite{sigma_x1}.


One surprising result is that the environmental noise believed to destroy quantum information, plays an important role in the generation of chaos in quantum systems. The reasons can be concluded as: (1) the environmental noise keeps absorbing photons from the intracavity, bringing about energy dissipation and decoherence. This creates a basin of attraction in the generalized quantum orbit, a necessary condition for chaos.
(2) More importantly, noise the memory effect in the quantum mechanical mode, such that its motion is dominated by the classical chaotic input.

\begin{center}
\begin{figure}[]
\centering
\includegraphics[width=3.4in]{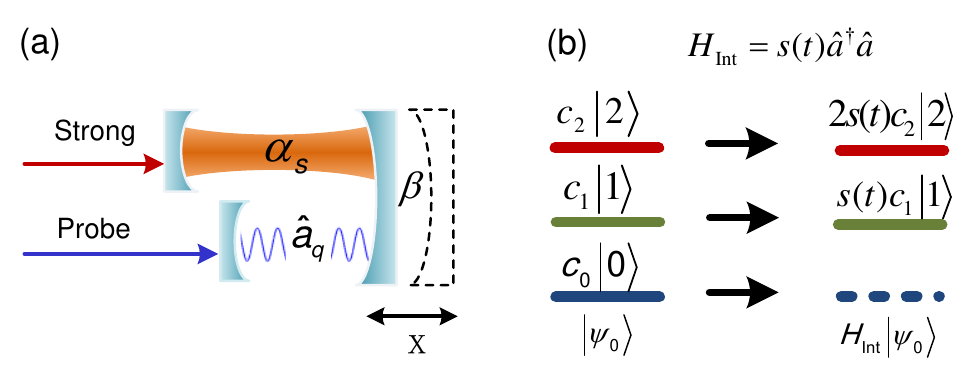}
\caption{(color online) (a) Setup for importing chaos into a quantum cavity (configuration II). A quantum cavity $\hat{a}_q$ is coupled to a classical chaotic optomechanical system via the mechanical mode $\beta$. (b) Energy level distribution of the quantum cavity without (left part) and with (right part) coupling to the classical signal $s(t)$.}
\label{fig04}
\end{figure}
\end{center}

\textit{Configuration II.} The setup is given in Fig.~\ref{fig04}, a quantum cavity $\hat{a}_q$ and a classical cavity $\alpha_s$ are connected to the same classical mechanical mode $\beta$. In this setup, chaos is supposed to be generated in the classical optomechanical resonator ($\alpha_s$, $\beta$), and then is imported into the quantum cavity $\hat{a}_q$ via the mechanical model $\beta$.

We begin with the classical part, i.e., the optomechanical resonator ($\alpha_s$, $\beta$). Its equations of motion are given by
\begin{subequations}\label{classical_eq1}
\begin{align}
\dot{\alpha}_s &=-i\Delta_s \alpha_s - \frac{\gamma_s}{2}
\alpha_s-2 i g_s \alpha_s x
+\varepsilon_s \;,\\
\dot{\beta} &= \left(-i \Omega - \frac{\Gamma}{2}\right) \beta - i g_s |\alpha_s|^2\;,\label{subeq:2}
\end{align}
\end{subequations}
where $\Delta_{s}$, $\gamma_s$, and $\varepsilon_s$ refer to the detuning frequency, damping rate, and driving strength of $\alpha_s$; while $\Omega$ and $\Gamma$ are the resonance frequency and damping rate of $\beta$; and their coupling strength is $g_s$.
Here, the mechanical mode $\beta$ is coupled to both the classical and quantum cavities ($\alpha_s$ and $\hat{a}_q$), enabling this classical-to-quantum chaos transfer.

Now, we concentrate our attention on the quantum part, i.e., the quantum cavity $\hat{a}_q$. We first consider the total system Hamiltonian
\begin{equation}\label{H1}
H_{\rm eff}=\Delta_q \hat{a}_q^{\dag}\hat{a}_q + s(t) \hat{a}_q^{\dag}\hat{a}_q,
\end{equation}
where $\Delta_q$ is the detuning frequency of $\hat{a}_q$. Also, $s(t)=2 g_q x(t)$ is the classical input, where $x(t)=(\beta+\beta^*)/2$ is the mechanical displacement
 and $g_q$ is the coupling strength between $\beta_s$ and $\hat{a}_q$.
Accordingly, the master equation can be written as
\begin{equation}\label{master_eq_1}
\dot{\rho}=i[\rho,{H}_{\rm eff}(t)] + \gamma_q \mathcal{D}[\hat{a}_q]\rho,
\end{equation}
where $\rho$ represents the density matrix, and $\mathcal{D}[\hat{a}_q]\rho=\hat{a}_q\rho\hat{a}_q^{\dag}-(\hat{a}_q^{\dag}\hat{a}_q\rho + \rho \hat{a}_q^{\dag}\hat{a}_q)/2$ is the Liouvillian in the Lindblad form for $\hat{a}_q$.

In the weak-driving limit, we limit the photon number of the quantum cavity $\hat{a}_q$ to be 3. The density matrix is then given by $\rho=\sum_{j,k=0}^{2}\lambda_{j,k}|j\rangle \langle k|$ in the Fock state representation. Recall that in configuration II, chaos in the quantum cavity $\hat{a}_q$ is encoded into the coefficients $\lambda_{jj}$ of the diagonal elements of the density matrix $\rho$ ($j=0,1,2$). In order to decode this chaotic motion, we reconstruct the generalized quantum orbit of the quantum cavity $\hat{a}_q$ from $N$ time-delayed coordinates $\lambda_{1,1}(\tau)$, $\lambda_{1,1}(2\tau)$, ..., $\lambda_{1,1}(N\tau)$ with the delayed time $\tau$ ($\tau=4$~{\rm ns}).

As shown in Fig.~\ref{fig04}, a chaotic attractor appears in the generalized quantum orbit of the quantum optical cavity $\hat{a}_q$. The classical chaotic input $s(t)$ introduces a chaotic transition
 or jump among different energy levels. Here, as an indicator of chaos, the largest Lyapunov exponent of $\lambda_{1,1}(t)$ is positive.
Experimentally, $\rho_{1,1}(t)$ can be readout by recording the mean photon number $\bar{n}_a(t)$, i.e., $\rho_{1,1}(t) \approx \bar{n}_a$.





We further explore the relationship between the classical and quantum components.
Both in setups Figs.~2(a) and 4(a), we find that the classical part works as a controller, dominating the motions of the quantum object.
When the classical system is prepared to periodic, multi-periodic, and chaotic regimes, the quantum object is also modulated to the corresponding states (Supplementary Figs.~2, 4, and 6).
\begin{center}
\begin{figure}[th]
\centering
\includegraphics[width=3.3in]{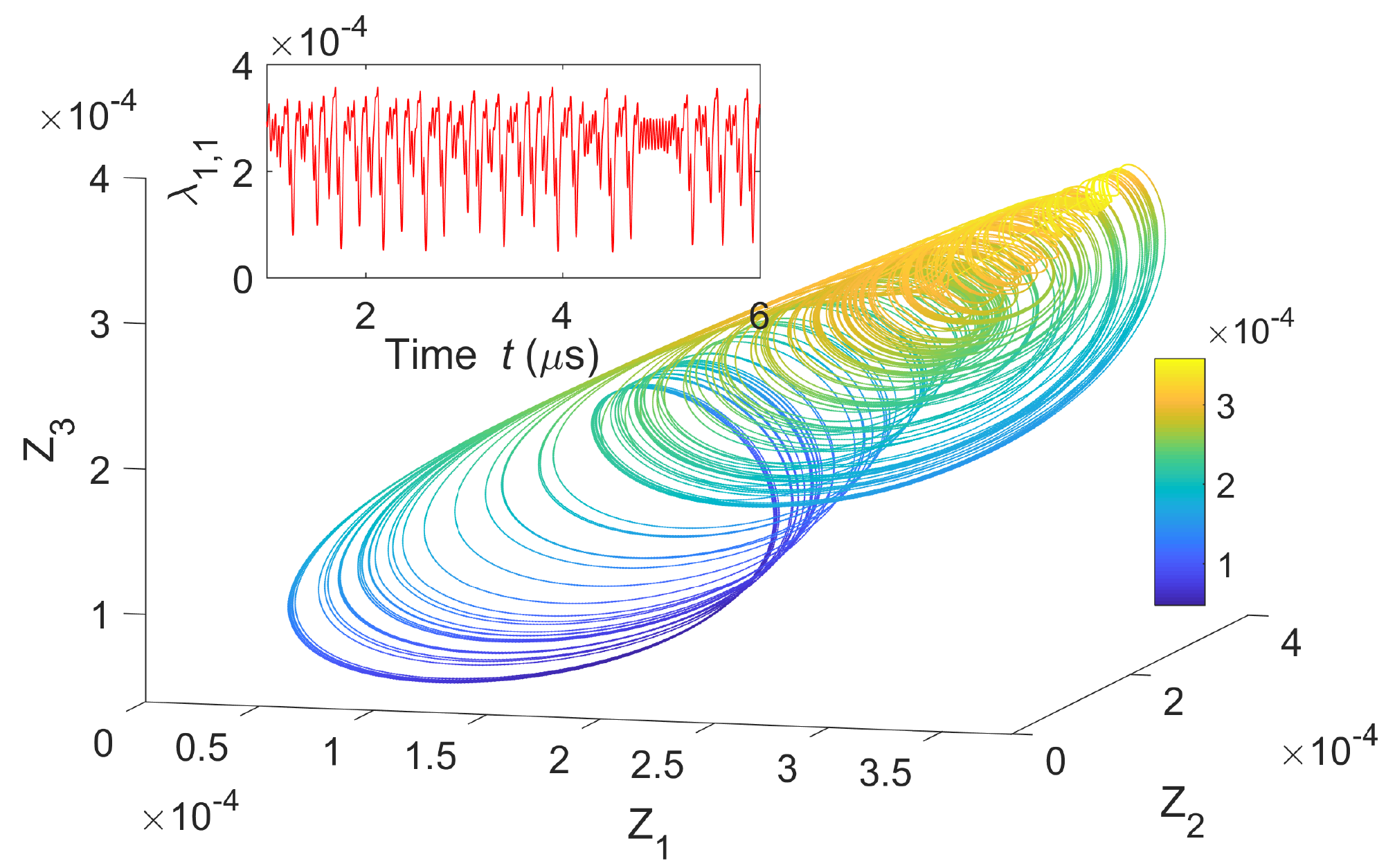}
\caption{(color online) Generalized quantum orbit of the quantum cavity $\hat{a}_q$ reconstructed from $\lambda_{1,1}(t)$, where $\bar{n}_a \approx \lambda_{1,1}(t)$. Here, $Z_1$, $Z_2$, and $Z_3$ refer to the three coordinates of the 4-dimensional phase space and the forth one $Z_4$ is presented by different colors. The parameters are:
$\Delta_s/\Omega=-1$, $g_s/\Omega=g_q/\Omega=0.1$, $\gamma_s/\Omega=\gamma_q/\Omega=1$, $\Delta_q/\Omega=1$, $\varepsilon_s/\Omega=4.33$, $\varepsilon_q/\Omega=0.01$,
$\Gamma/\Omega=10^{-3}$, and $\Omega/2\pi=0.1$~{\rm GHz}.}
\label{fig05}
\end{figure}
\end{center}


\textit{Conclusions and discussions.} ---
This work paves a way to study quantum chaotic systems in the framework of chaos theory, i.e., by extracting the information of a quantum chaotic system and decoding it as a chaotic attractor.
Moreover, we import classical chaos into a quantum system, providing a feasible method for designing quantum chaotic systems.
The questions raised in Introduction can be summarized as: (1) Based on configurations I or II, quantum chaotic motions can be strictly defined~\cite{Zurek1995QuantumCA,Slomczynski1994} in the framework of  chaos theory, e.g., Devaney's Definition~\cite{Devaney}. (2) A quantum system is linear {when} it is isolated and non-time-delayed. However, nonlinear regimes can occur in a quantum system that is coupled to a time-delay term (kicked rotor), or a classical signal as discussed in this paper.

Below are several outlooks for future studies.
(1) The "Generalized quantum orbits" proposed in this work can also be applied to the quantum chaos generated by the Kicked rotor, the quantized baker's map, and the Bunimovich
stadium billiard.
(2) Configurations I and II potentially enable the compatibility between quantum mechanics and classical dynamical theory. Therefore, they could work as the starting point of some other complex phenomena in quantum systems, e.g., quantum synchronization.
(3) We offer an easily controlled method to adjust the motion of a quantum object, which can be used to create nonlinear quantum signals, e.g., nonlinear photonic quantum gate~\cite{Gullans2013,Cox2014,Hendry2010} and quantum mechanical memory~\cite{Leijssen2017,Caspani2017}.
(4) Note that noise, known as the quantum information destroyer, also acts as a quantum chaos creator. It would also be interesting to ask how different types of noise contribute to the generation of complex quantum behavior.

\begin{acknowledgments}
NY would like to thank Yuping Huang, Quanzhen Ding, and Clemens Gneiting for useful discussions.
F.N. is supported in part by: NTT Research, Army Research Office (ARO) (Grant No. W911NF-18-1-0358), 
Japan Science and Technology Agency (JST) (via the CREST Grant No. JPMJCR1676), Japan Society for the Promotion of Science (JSPS) (via the KAKENHI Grant Number JP20H00134, JSPS-RFBR Grant No. 17-52-50023),  and Grant No. FQXi-IAF19-06 from the Foundational Questions Institute Fund (FQXi), a donor advised fund of the Silicon Valley Community Foundation. 
Y.L. is supported by the National Natural Science Foundation of China (NSFC) (Grants No. 91736106, 11674390, 91836302).
\end{acknowledgments}

\bibliography{references1}

\end{document}